**Optimized, Unequal Pulse Spacing in Multiple Echo Sequences Improves Refocusing**

Elizabeth R. Jenista, Ashley M. Stokes, Rosa Tamara Branca and Warren S. Warren

Department of Chemistry and Center for Molecular and Biomolecular Imaging, Duke University

Abstract: A recent quantum computing paper (G. S. Uhrig, *Phys Rev Lett* 98 100504 (2007)) analytically derived optimal pulse spacings for a multiple spin echo sequence designed to remove decoherence in a two level system coupled to a bath. The spacings in what has been called a "UDD sequence" differ dramatically from the conventional, equal pulse spacing of a Carr-Purcell-Meiboom-Gill (CPMG) multiple spin echo sequence. The UDD sequence was derived for a model that is unrelated to magnetic resonance, but was recently shown theoretically to be more general. Here we show that the UDD sequence has theoretical advantages for magnetic resonance imaging of structured materials such as tissue, where diffusion in compartmentalized and microstructured environments leads to fluctuating fields on a range of different timescales. We also show experimentally, both in excised tissue and in a live mouse tumor model, that optimal UDD sequences produce different $T_2$-weighted contrast than do CPMG sequences with the same number of pulses and total delay, with substantial enhancements in most regions.

Corresponding author: Warren S. Warren, Duke University, Box 90346, Durham, NC 27708, phone (919)660-1508, fax: (919)287-2454, warren.warren@duke.edu.

**Introduction**

Quantum computing--the manipulation of a quantum mechanical system to do information processing--has attracted considerable attention since the mid-1990s, largely triggered by Shor's proposed algorithm for finding prime factors in polynomial instead of exponential time [1]. One of the earliest proposed implementations was solution phase NMR [2], due to the long coherence times (leading to readily resolved resonances) and simple spin Hamiltonian. Despite theoretical objections that solution NMR quantum computers could never be scalable to a useful number of "qubits"[3], this field saw substantial experimental research efforts aimed at developing implementations of quantum computing algorithms, with the high water mark probably being the demonstration that a seven-spin system could be made to factor the number 15 [4]. In more recent years, experimental interest has largely shifted to systems such as trapped ions and quantum dots that in principle might be scalable. However, the theoretical framework has led to new insights into the nature of coherence, correlation, and entanglement, and to new ways to think about pulse sequences.

Here we present what might be the first case of this framework enabling magnetic resonance applications, in this case extending the coherence lifetime in MR tissue imaging. We start from a recently published result[5] which predicted a specific optimal, highly *nonuniform* spacing for multiple echo sequences used to prevent decoherence (as opposed to the equal spacing used in virtually all MR applications). Their model for relaxation was not directly relevant to magnetic resonance. Here we adapt this work to show theoretically that their predicted spacing can significantly strengthen magnetic resonance signals in the limit normally encountered in structured samples, and we verify this experimentally in tissue samples and live animals. This also suggests new possibilities for enhancing contrast and for understanding tissue dynamics.

**Theory of relaxation during multiple spin echo sequences**

For the simplest ensembles of two-level systems, such as those found in solution NMR, the distinction between a "homogeneous" $T_2$ and "inhomogeneous" $T_2^*$ lifetime is easy to make: local variations in the magnetic field can be treated as constant over the time between pulses, giving a static contribution to the spin Hamiltonian of the form $H = \hbar \Delta\omega(\vec{r}) I_z$. In this limit (ignoring all pulse imperfections for the moment), a simple spin echo gives the same signal as a multiple echo sequence (Figure 1a) of the same total length. Such multiple echo sequence are traditionally viewed in an interaction representation where the π pulses rotate the Hamiltonian instead of the initial magnetization, thus creating what is called a "toggling frame Hamiltonian"[6] $\tilde{H}$. In this case, $\tilde{H}$ changes sign after each π pulse:

$$\tilde{H} = \begin{cases} \hbar\Delta\omega(\vec{r})I_z \text{ during } t_1, t_3, t_5, \ldots \\ -\hbar\Delta\omega(\vec{r})I_z \text{ during } t_2, t_4, t_6, \ldots \end{cases}$$

$$\equiv \hbar\Delta\omega(\vec{r})S(t)I_z, \text{ where } S(t) = \begin{cases} 1 \text{ during } t_1, t_3, t_5, \ldots \\ -1 \text{ during } t_2, t_4, t_6, \ldots \end{cases} \tag{1}$$

In equation (1) we introduce *S(t)*, the modulation of the Hamiltonian created by the pulses (which is also the function graphed in Figure 1b). The only spin operators ever present are $\pm I_z$, which makes it trivial to calculate the effects of the pulse sequence. As long as the sum of the even delays is equal to the sum of the odd delays, the average of *S(t)* is zero, and the effect of resonance offset is removed (Figure 1(b)). In conventional notation[6], the "average Hamiltonian" $\overline{H} = \int \tilde{H}(t) dt = 0$.

Making *S(t)* average to zero imposes only a single constraint on the delays, while leaving many degrees of freedom to optimize them. For example, diffusion in a constant magnetic field gradient imposes an additional decay ($\exp(-\alpha t^3)$) on the magnetization[7,8]. It is then trivial to show that for a fixed total echo time and number of pulses, the best positioning is the Carr-Purcell-Meiboom-Gill (CPMG) [7,9] sequence shown in Figure 2a: equal delays *T/n* between the *n* π pulses and delays of *T/2n* before the first and after the last π pulse if *n* is even, or all delays equal to *T/(n+1)* if *n* is odd. Some refinements have been made to CPMG experiments over the decades since they were introduced (including the introduction of composite [10] or shaped adiabatic pulses [11], or of crusher gradients around each π pulse), but the basic structure of equal pulse spacing has remained uncontested.

In recent years, numerous experiments have shown that in tissue, a 10-100 ms CPMG sequence often gives a higher signal intensity than does a spin echo sequence of the same total duration even without an externally applied gradient[11]. This arises from at least two effects. Susceptibility variations within even the smallest image voxels combine with diffusion to generate magnetic field fluctuations; in addition, magnetization transfer and migration between different compartments modulate the resonance frequency. Both of these effects create a time- and position-dependent resonance offset $\Delta\omega(\vec{r},t)$ with spectral density over a range of frequencies:

$$H = \hbar\Delta\omega(\vec{r},t)I_z = \hbar\left(\int G(\vec{r},\omega)e^{-i\omega t}d\omega\right)I_z \qquad (2)$$

(In what follows for simplicity we will drop the explicit position dependence, thus writing *Δω(t)* or *G(ω)*). From equation (2) different groups of spins accumulate different extra random phase shifts, and the echo disappears; in practice, spin echo relaxation times of water in tissue at high fields can be a factor of 100 or more shorter than in a test tube.

This is not the same physical model as motion in a constant gradient, but it remains intuitively reasonable to avoid long delays between pulses, hence CPMG would seem to be the best choice. Remarkably, a recent theoretical quantum computing paper [5] shows this is not in general correct. That paper showed that under specific circumstances, the optimum timing for $n$ pi pulses is not the equidistant one; in our notation (centering the sequence at $t=0$), the $j^{th}$ pulse should be located at time

$$\delta_j = T \{\sin^2(\pi j/(2n+2))-0.5\}. \qquad (3)$$

This set of timings is now commonly cited in the quantum computing community as the "Ührig dynamic decoupling" or UDD sequence and is illustrated in Figure 2b for the sixteen-pulse version. The sequence is symmetric, with longer delays near the center. While the relaxation model in that paper is not directly relevant to magnetic resonance, the simplicity of the results led to speculation [12] and later proof [13] that this was a universal result for a broad class of models of relaxation in two-level systems. This work has also been extended theoretically [14] to more complex, concatenated pulse sequences, and lifetime extension has been experimentally demonstrated for ion coherences in Penning traps [15].

These papers are written in language which is unfamiliar to the magnetic resonance community. Here we rephrase their results, adapt them to the average Hamiltonian formulation, and slightly alter the model to make it more appropriate for magnetic resonance. Consider a single frequency component in equation (2), $\hbar G(\omega)e^{i\omega t}I_z$. A multiple echo sequence will create a toggling frame Hamiltonian and average Hamiltonian of

$$\widetilde{H} = \hbar G(\omega)S(t)e^{-i\omega t}I_z; \overline{H} = \int \hbar G(\omega)S(t)e^{-i\omega t}I_z dt = \hbar G(\omega)S(\omega)I_z \qquad (4)$$

Figure 1c shows how $\tilde{H}$ is modulated, starting with a single sine wave component. Even if the sine wave itself were to go through an integral number of cycles during the pulse sequence, the average $\overline{H}$ would not generally be zero. Thus, a multiple echo sequence converts resonance frequency fluctuations at frequency ω into an average frequency shift which depends on the phase of the sine wave. Since G(ω) is expected to vary in phase across the sample, this ends up causing position-dependent frequency shifts, hence signal dephasing proportional to $|G(\omega)|^2|S(\omega)|^2$.

Changing the delays in a multiple echo sequence changes $S(t)$ and $S(\omega)$, but Parseval's theorem shows that:

$$\int_{-\infty}^{\infty} |S(\omega)|^2 d\omega = \int_{-T/2}^{T/2} |S(t)|^2 dt = \int_{-T/2}^{T/2} (1) dt = T \tag{4}$$

Thus, if the resonance frequency fluctuations were uniform in frequency, changing the delays makes no difference at all-any multiple pulse echo sequence (or no pulses at all, $S(t)=1$) would perform exactly the same. However, since spin echoes do produce refocusing in tissue, and CPMG sequences produce additional refocusing, we can infer that low frequency components (relative to 1/T) play an important role in the relaxation dynamics. This is also physically reasonable-for example, apparent diffusion coefficients in water correspond to motion over cellular distances (and hence exposure to susceptibility differences) in tens of milliseconds.

As noted above, CPMG or any other sequence with $t_1+t_3+t_5...=t_2+t_4+t_6...$ makes the average value of $S(t)$ (=$S(\omega=0)$) vanish, and thus refocuses static resonance frequency variations. For any sequence symmetric about $t=0$ (including CPMG with an even number of pulses), all odd derivatives of $S(\omega)$ vanish at the origin. The key result of reference [5] is that the specific pulse placements in the UDD sequence make all of the first *(n-1)* derivatives $d^m S(\omega)/d\omega^m$ vanish at *ω=0* (equivalently, the first error term signal scales as the *(n+1)* power of the sequence length). This is a truly remarkable result; in average Hamiltonian theory, methods have been derived to create an arbitrarily high power scaling with sequence length (reference [16] presents an early example), but they generally require an exponentially increasing number of pulses. Thus, the suppression of relaxation effects at moderately low frequencies is optimally effective.

Figure 3 shows explicit calculations for a sixteen pulse UDD sequence illustrating this point. Figure 3a shows that the first *15* derivatives at the origin vanish for the UDD sequence; for CPMG, even the second derivative is nonzero. The consequence of this suppression for different frequencies is shown in Figure 3b. For almost all frequencies up to $\omega=25/T$, $S(\omega)$ is smaller for UDD than for CPMG and hence those frequencies are less effective in dephasing. Incidentally, the zeroes on the graph for CPMG at $\omega=2\pi/T$, $4\pi/T$…are because of the periodic structure, but $\omega=16\pi/T$ is very large: at that frequency, $S(\omega)$ goes through a $\pi$ phase shift during the delay between the pulses, and the sine wave is effectively rectified.

Reference [14] suggests that there would be applications to "high-precision NMR where narrow linewidths are a prerequisite," but this would appear to be unpromising. Multiple spin echo sequences remove or distort scalar couplings, and in any event the linewidth improvements are likely to be miniscule as resonance frequency fluctuations on the relevant timescale (seconds) are very small in an unstructured sample. However, $T_2$ extension is of great value in a variety of tissue imaging applications. For example, the most common intermolecular multiple-quantum coherence (iMQC) signals grow in linearly with time[17], and in most such applications, the maximum signal strength is directly proportional to $T_2$. Pairs of hyperpolarized spins can be prepared in the "singlet state" $\alpha\beta-\beta\alpha$ to increase their lifetime[18,19], but if the two spins are inequivalent, multiple echo sequences are used to prevent interconversion to $\alpha\beta+\beta\alpha$[18] and even equivalent spins[19] might see a lifetime extension. More generally, different tissue microstructures would generate different resonance frequency fluctuations $G(\omega)$ that would not be expected to respond identically to the UDD sequence, so it could provide a new source of contrast reflecting subvoxel information that is not equivalent to any other existing sequences.

**Experimental Methods**

To test the hypothesis that UDD sequences provide additional refocusing, we decided to compare imaging sequences on a postmortem mouse (Figures 4-5) and in vivo in a tumor bearing mouse (Figure 6). The mouse was placed in a supine position, and axis slices through the abdomen or through the tumor were selected. MRI data were acquired on a Bruker 7.05T ($^1$H: 300.5 MHz). In all cases we compared the contrast from three sequences: UDD, CPMG, and an "anti-UDD" sequence (Figure 1c) with the same total delay and number of pulses, hence the same dissipated power. Pulse positions for the UDD sequences were calculated using equation (3). The anti-UDD sequence was done as a control experiment. This sequence has exactly the same set of delays as UDD, and it still has $S(\omega=0)=0$ (e.g., $t_1+t_3+t_5\ldots=t_2+t_4+t_6\ldots$) so the echo timing is correct; however, the intervals are rearranged to put the longer delays up front, and the shorter delays toward the end of the sequence. This will make even the first derivative $dS(\omega)/d\omega \neq 0$, and by the arguments above should make the refocusing performance worse.

For the post mortem experiments in Figures 4-5, 2 and 1.267 ms Hermite pulses were used for excitation and refocusing, respectively. No phase cycle was performed and the images were all acquired with NA=1. The echo time was varied between 40 and 80 ms. An 8mm axial slice was then selected through the abdomen with a 8 cm FOV and a 256 x 256 matrix size. All the 180 degree pulses in each sequence were flanked by crusher gradients.

For the comparison with a $T_2$ map (Figure 6) and the in vivo experiment (Figure 7), 0.5 ms Hermite pulses were chosen for both excitation and the 8 refocusing pulses. A 2mm axial slice was selected across the tumor. The total echo time for the spin echo, CPMG, and UDD was chosen equal to 120ms for the in vivo experiment, 100 ms for the $T_2$ map, with a total of 8 pi pulses for CPMG and UDD. Other imaging parameters were TR=2s and 3 cm FOV.

All images were processed using ImageJ (NIH, Bethesda, MD) and MATLAB software (Mathworks, Inc.).

**Results and Discussion**

Figure 4-5 compares CPMG, UDD and anti-UDD sequences for a postmortem mouse that had recently been thawed (giving a significant amount of free water, indicated by the arrows). In general, the CPMG sequence highlights the free water deposits in the subcutaneous layers; the UDD sequence highlights the tissue architecture; and the anti-UDD sequence give weaker signals overall. Differences are much more evident with more pulses (which increases the number of zeroed derivatives in the UDD sequence) and with longer echo times (which reduces the effects of duty cycle).

In most cases, the UDD sequence provides more signal in bulk tissue than both the CPMG and anti-UDD sequences (Figure 5, signal comparisons in Table 1). In particular, the difference images in Figure 5 show that that the UDD sequence provides better refocusing of the tissue signal, and the anti-UDD sequence does not have as much signal in bulk tissue as the CPMG or UDD sequence; for the long echo time (80 ms) sequence with 16 pulses, the UDD:CPMG:anti-UDD signal ratios are 1.7: 1: 0.34 in the indicated tissue ROI. However, the improvement provided by UDD is not just a uniform factor, and in fact the free water refocuses better with CPMG. Since the local structure is different in those regions, differences are to be expected. Figure 6 provides a direct comparison (different sample from Figures 4-5) of a $T_2$ map with an image of the difference between UDD and CPMG; from the figure it is clear that on average, regions with moderately short $T_2$ show better UDD signals than do the longest $T_2$ regions, but it is also clear that there is no simple relation between $T_2$ and UDD-CPMG. This implies the contrast in the latter image is not simply equivalent to what could have been inferred from simple $T_2$ weighting.

Figure 7 shows an in vivo application on a nude mouse bearing a human prostate tumor (the bright regions, showing a longer $T_2$ value, correspond to the necrotic area of the tumor, whose size exceeds 4 cubic cm). A trend similar to the post mortem experiment was seen in this in vivo study: most of the bulk tissue gives more signal with the UDD sequence, but the bright regions with free water do not. Understanding these trends (or, indeed, proving them to be universal) will require significant additional studies; however, since the UDD sequence serves as a very sharp high-pass filter for frequency fluctuations, careful variation of the overall sequence length could provide a very effective mapping method for $G(\omega)$, and could reasonably be expected to provide a wealth of structural information that is not accessible by conventional MR methods.

The analysis presented here so far has been idealized. A quantitative comparison of expected signal intensity between, for example, a spin echo and a CPMG sequence would require careful attention to pulse flip angles, effects of diffusion during crusher gradients, effects of finite pulse length, and dephasing effects due to scalar couplings in the non-water components of the tissue. However, comparisons between UDD, CPMG and anti-UDD sequences are much easier to make. The effects of imperfect flip angles (which, in the presence of crusher gradients, attenuate the echo), signal attenuation due to diffusion itself in the gradients, and finite duty cycle effects will be identical for UDD, CPMG, and anti-UDD. Scalar couplings are easily shown to be least important when the spacing is equal, as in CPMG. Thus nonidealities cannot explain the very substantial differences we see in tissue between these sequences.

In summary, we have shown that a nonintuitive, unequal spacing of pulses in a multiple echo sequence, as derived theoretically by Uhrig for decoherence reduction in quantum computing, gives improved performance in refocusing for some tissue types and also tends to suppress the (usually uninteresting) free water signals. We have not proven that the UDD sequence is a global optimum for magnetic resonance imaging; in fact, it is likely that different tissue types, with different spectral densities of the field fluctuations, will have different optimum solutions. However, the surprising result is that CPMG is clearly far from the optimum. There are many potential applications for such sequences, and other obvious directions for further improvements can be explored (e.g., phase shifts in the individual pulses, corrections for finite pulse widths, shaped rf pulses). Finally, while we have focused here on dephasing, predicted extensions in the quantum computing literature to $T_1$ lifetime improvements[14] might also be of value for $T_1$-weighted images.

This work was supported by NIH grant EB 02122.

Table 1:

|  | Free water | Tissue |
|---|---|---|
| **UDD/CPMG, 8 pulse, TE=20 ms** | **1.03** | **1.41** |
| Anti-UDD/CPMG, 8 pulse, TE=20 ms | 0.69 | 1.35 |
| UDD/Anti-UDD, 8 pulse, TE=20 ms | 1.50 | 1.05 |
| **UDD/CPMG 16 pulse, TE=40 ms** | **0.40** | **1.22** |
| Anti-UDD/CPMG 16 pulse, TE=40 ms | 0.27 | 0.93 |
| UDD/Anti-UDD, 16 pulse, TE=40 ms | 1.50 | 1.31 |
| **UDD/CPMG, 16 pulse, TE=80 ms** | **0.52** | **1.71** |
| Anti-UDD/CPMG, 16 pulse, TE=80 ms | 0.32 | 0.59 |
| UDD/Anti-UDD, 16 pulse, TE=80 ms | 1.65 | 2.88 |

Table 1: Comparison of the signal strength from the UDD, anti-UDD and CPMG sequences. Free water and tissue ROIs were selected as shown in figure 4. Note that at the longer echo times with 16 pulses, the CPMG sequence refocuses the free water very well, while the UDD sequence has significantly improved SNR in the tissue ROI. In addition, the anti-UDD sequence

significantly underperforms both the UDD and CPMG sequence in the tissue ROI for the 16 pulse sequence with TE=80ms.

**Figure captions**

Figure 1a. Generic multiple echo sequence. All pulses are π pulses with the same phase; the position of the $i^{th}$ pulse is $\delta_i$; and the delay just before the $i^{th}$ pulse is $t_i$. Figure 1b. Static resonance frequency variations are rotated by the π pulses, giving a "toggling frame Hamiltonian" $\tilde{H}$ proportional to $\pm I_z$. As long as $t_1+t_3+t_5...=t_2+t_4+t_6...$, the effect cancels since $\tilde{H}$ averages to zero. Figure 1c. A time-varying frequency fluctuation (sin(ωt)) is altered by the pulse sequence, but in general $\tilde{H}$ is not averaged to zero. The UDD sequence in Figure 2 does the best possible job of cancelling low-frequency fluctuations.

Figure 2: 16 pulse CPMG, UDD, and anti-UDD pulse sequences. For example, for an 80 ms total echo time, the CPMG sequence has a uniform delay between the p pulses of 2.5 ms, except for the first and last delay, which are 1.25 ms each. The delays in the UDD sequence are .681, 2.02, 3.2902, 4.4483, 5.455, 6.2759, 6.883, 7.2558, 7.3814, 7.2558, 6.883, 6.2759, 5.455, 4.4483, 3.2902, 2.02, and .681 ms. The "anti-UDD" sequence (see text) is a control sequence with the same delays as the UDD sequence, arranged in different order.

Figure 3a. The pulse spacing in an *n*-pulse UDD sequence suppresses the effects of resonance frequency fluctuations at low frequency, making the first (*n-1*) derivatives around ω=0 vanish. This figure compares the even derivatives of the 16 pulse version to a CPMG sequence (which has a nonvanishing second derivative). Odd derivatives vanish for both. Figure 3b. Explicit comparison of the efficiency of frequency fluctuations (for example, from diffusion in a structured sample with susceptibility differences) in causing dephasing. This figure compares a 16 pulse UDD sequence and a 16 pulse CPMG sequence of the same duration. Note that for low frequency modulation, UDD vastly outperforms CPMG in inhibiting relaxation.

Figure 4: Spin echo image of the post-mortem mouse used in the experiments in Figure 5. Arrows indicate locations in the mouse where there is excess free water, since the mouse was frozen and thawed. The boxes in red show the ROIs used for the analysis of signal strength in table 1.

Figure 5: Comparison of the UDD, CPMG and anti-UDD pulse sequences on the postmortem mouse in Figure 4. The effect of the UDD sequence is most apparent at the longer echo times, with larger numbers of pulses.

Figure 6: Left: $T_2$ map on a postmortem mouse (different from the one in Figures 4-5) obtained by fitting spin echo data from 10-160 ms. Right: difference image between 8-pulse UDD and 8-pulse CPMG (sequence length 80 ms). Regions with moderate $T_2$ benefit more from the UDD sequence than do the long $T_2$ regions.

Figure 7. In vivo axial images of tumor tissue obtained with 8 pulse UDD and CPMG sequences and with a spin echo sequence (120 ms total echo time). The tumor tissue appears highly inhomogenous with several necrotic areas with a higher signal intensity. Differences between UDD, CPMG and spin echo are typically $\pm 25\%$.

Figure 1

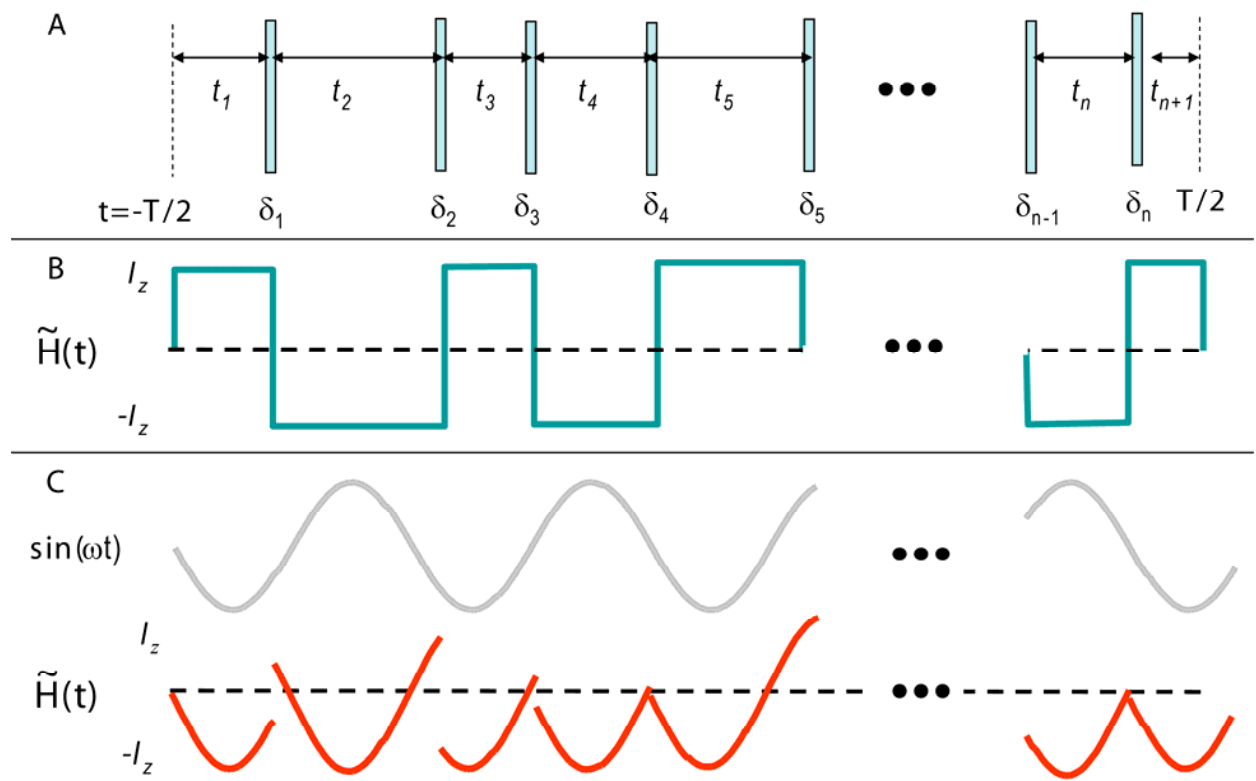

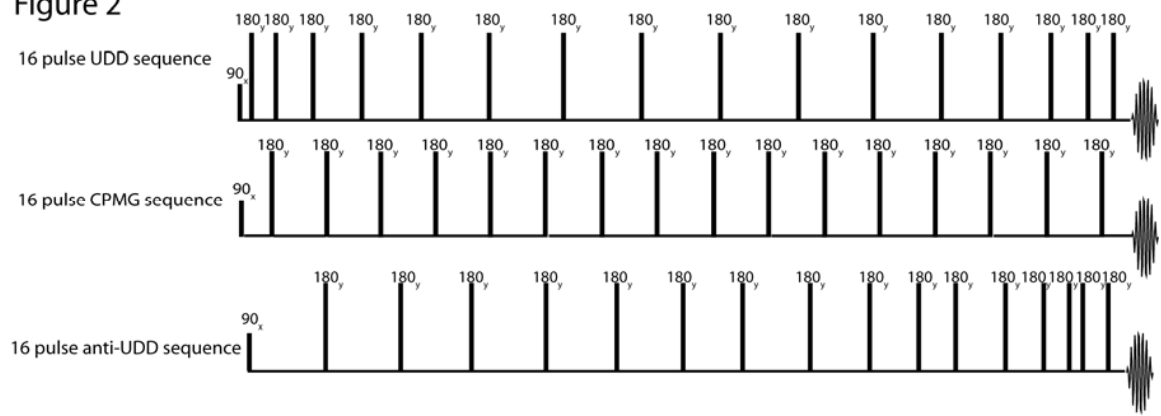

Figure 2

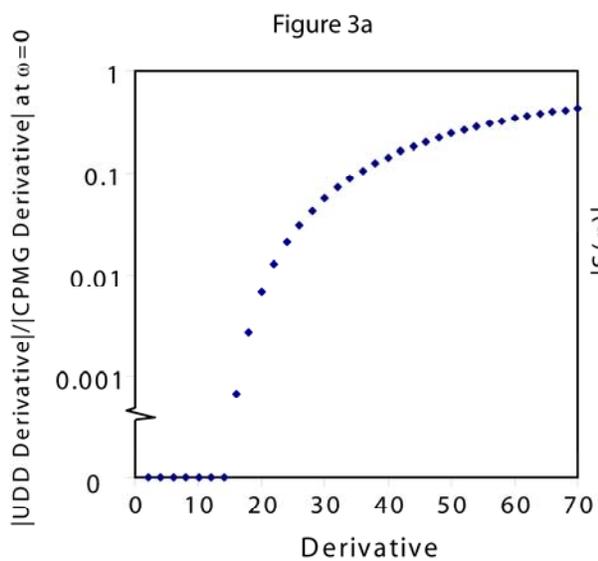

Figure 3a

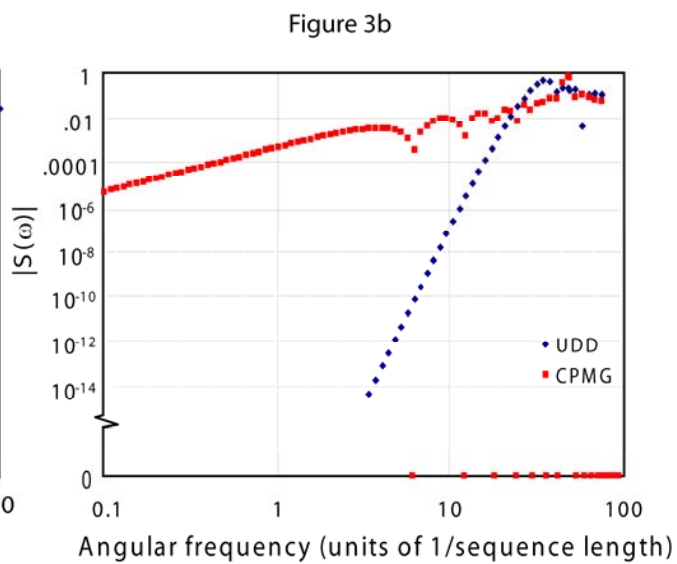

Figure 3b

Figure 4

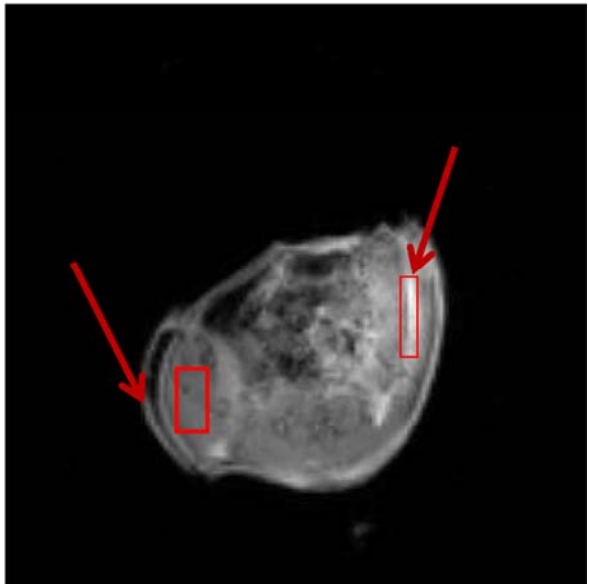

Figure 5

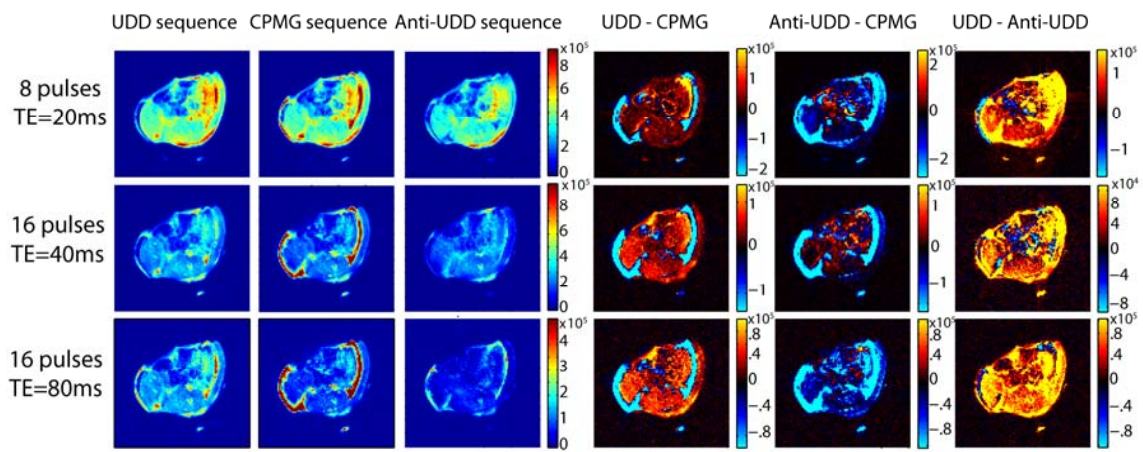

Figure 6

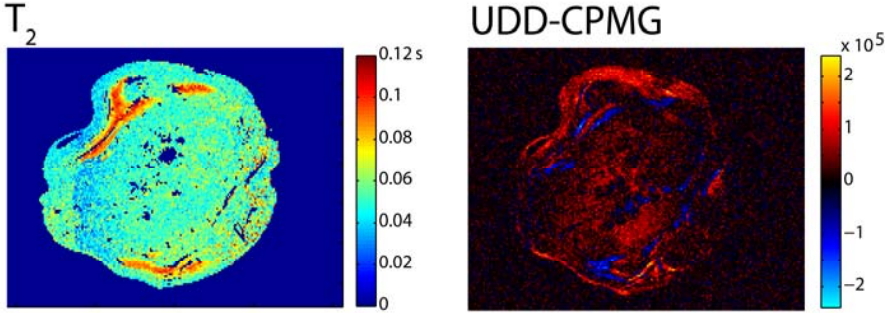

Figure 7

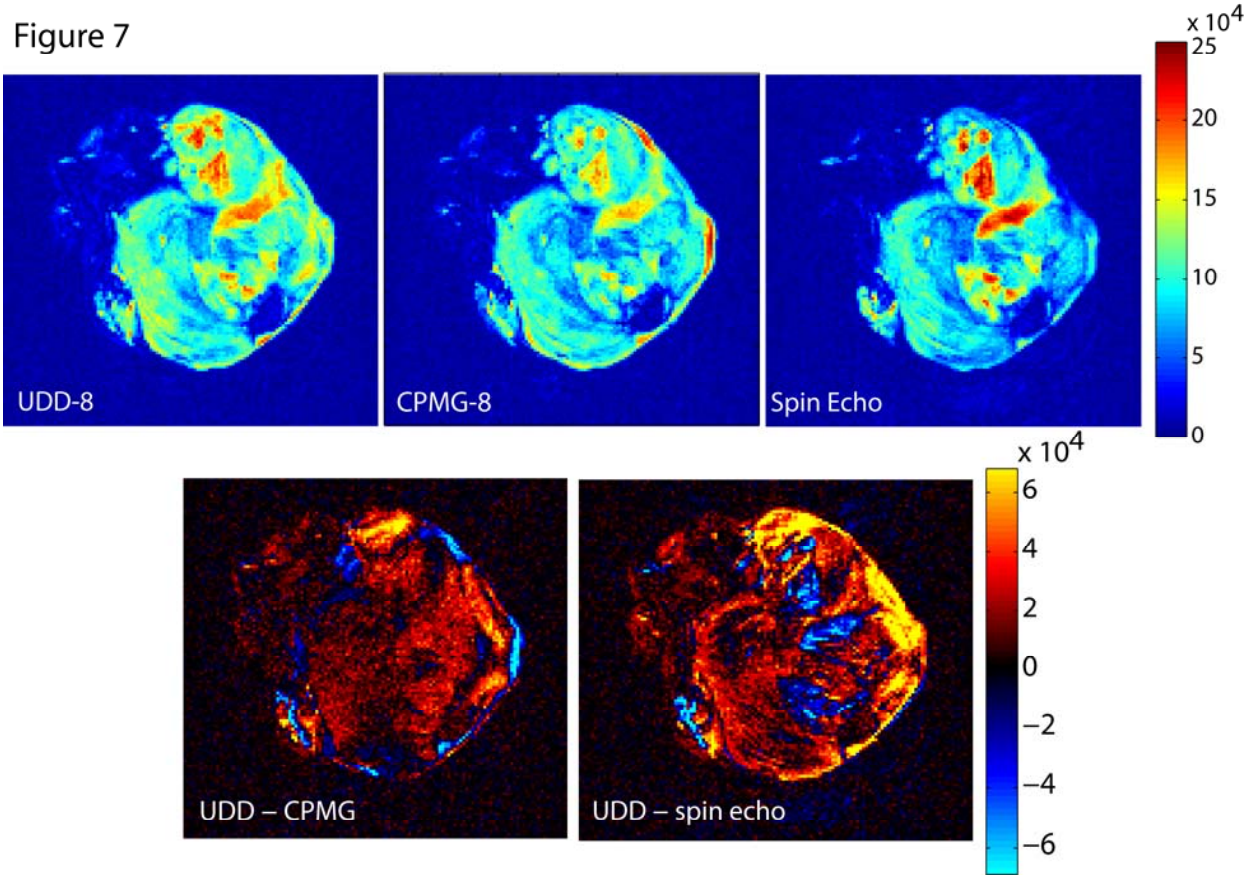